\documentclass[hyper,12pt,letterpaper]{JHEP}
\usepackage{mathrsfs}
\usepackage{amsmath}
\usepackage{amsfonts}
\usepackage{amssymb}
\usepackage{graphicx}

\DeclareSymbolFont{AMSa}{U}{msa}{m}{n}
\DeclareSymbolFont{AMSb}{U}{msb}{m}{n}
\let\Box\relax
\DeclareMathSymbol{\Box}{\mathord}{AMSa}{"03}

\def\bc{\begin{center}}
\def\nno{\nonumber}
\def\ec{\end{center}}
\def\be{\begin{eqnarray}}
\def\ee{\end{eqnarray}}

\def\al{\alpha}
\def\ga{\gamma}

\def\eps{\epsilon}

\def\la{\lambda}

\def\Ga{\Gamma}
\def\Dl{\Delta}

\def\r{\partial}

\newcommand\rha{\rightarrow}

\title{Analytical Studies on Holographic Superconductors in Gauss-Bonnet Gravity}

\author{~Huai-Fan Li$^{~1,~2,~3}$,Rong-Gen Cai$^{~2}$,~Hai-Qing Zhang$^{~2}$\\
  $^{1}$Department of Physics and Institute of Theoretical
   Physics,\\
Shanxi Datong University, Datong 037009, China \\
  $^{2}$Key Laboratory of Frontiers in Theoretical Physics,\\
Institute of Theoretical Physics, Chinese Academy of Sciences,\\
   P.O. Box 2735, Beijing 100190, China\\
 $^{3}$Department of
Applied Physics, Xi' an Jiaotong University, \\ Xi' an 710049, China

{\tt E-mail:
\email{huaifan.li@stu.xjtu.edu.cn,cairg@itp.ac.cn,hqzhang@itp.ac.cn}}}

\abstract{ We use the variational method for the Sturm-Liouville
eigenvalue problem to analytically calculate some properties of
holographic superconductors with Gauss-Bonnet gravity in probe
limit. By studying the holographic p-wave and s-wave superconductors
in (3+1)-dimensional boundary field theories, it is found that near
the critical temperature, the critical exponent of the condensation
is $1/2$ which is the universal value in mean-field theory. We also
find that when Gauss-Bonnet coefficients grow bigger the operators
on the boundary field theory will be harder to condense. These are
in good agreement with the numerical results.}


\begin{document}


%

\section{\bf Introduction }

The AdS/CFT
correspondence~\cite{Maldacena:1997re,Gubser:1998bc,Witten:1998qj}
provides a theoretical method to understand strongly coupled field
theories. Recently, it has been proposed that the AdS/CFT
correspondence can also  be used to describe superconductor phase
transitions~\cite{Gubser:2008px,Hartnoll:2008vx}. Since the high
$T_c$ superconductors are shown to be in the strong coupling regime,
one expects that the holographic method could give some insights
into the pairing mechanism in the high $T_c$ superconductors.

There have been lots of works studying various holographic
superconductors, see review \cite{Hartnoll:2009sz,Herzog:2009xv}, in
which some effects such as scalar field mass,  external magnetic
field, and back reaction {\it etc.}, have been discussed. Among
those works, some universality is discovered. For instance, the
critical exponents for the condensation near the critical
temperature is found to be $1/2$ which is a universal value in
mean-field theory \cite{Bardeen:1957mv}. However, this value of
$1/2$ is mostly obtained by numerically solving the holographic
systems and then fitting the data. The frequently asked question is
how to calculate this $1/2$ from analytical study. In
\cite{Siopsis:2010uq}, the authors used the variational method for
the Sturm-Liouville (S-L) eigenvalue problem\footnote{For short, we
will call this method used by \cite{Siopsis:2010uq} to be ``S-L
method" in the following context.} to analytically calculate some
properties of the holographic superconductors in a (2+1)-dimensional
boundary field theory. They managed to obtain the critical exponent
$1/2$ by perturbing the system near the critical temperature.
Afterwards, the authors in \cite{Zeng:2010zn} extended this S-L
method to study the holographic s, p, d-wave superconductors in
3-dimensional boundary field theory. They found that this $1/2$ is a
universal value for these models.

In this paper, we extended the S-L method to study holographic
p-wave and s-wave superconductors in a (4+1)-dimensional bulk with
Gauss-Bonnet gravity. It is not trivial to extend this analytical
studies to (4+1)-dimensional case, because the dimension of
space-time will affect the boundary behavior of the fields which
coupled to gravity. And moreover it would further have an impact on
the conformal dimension of the dual operators. For instance, in
(2+1)-dimensional boundary field theory the charge density
$\rho\propto T^2$ where $T$ is the temperature; However, in
(3+1)-dimensional boundary field theory the relation is $\rho\propto
T^3$. This discrepancy would make the analysis a little different
from the one in \cite{Siopsis:2010uq}. We will see it explicitly in
the following context. Besides, we also calculated the ratio between
the critical temperature and the charge density for various
Gauss-Bonnet couplings. The ratio was greatly consistent with the
previous numerical results \cite{Gregory:2009fj,Cai:2010cv}. We also
found that the greater the Gauss-Bonnet coefficients were, the
harder the condensations to form, which was also consistent with the
previous conclusions \cite{Pan:2009xa,Cai:2010zm,Brihaye:2010mr}.
Although the conformal dimensions of the dual operators on boundary
theory are different from those in \cite{Siopsis:2010uq}, there
still exists a universal critical exponents $1/2$ as in the
mean-field theory. The ratios between the condensation values and
the critical temperatures matched the numerical results up to the
same order.

The paper is organized as follows: In Sec.\eqref{sect:gb} we briefly
review the basic facts of Ricci flat Gauss-Bonnet-AdS black holes.
In Sec.\eqref{sect:pwave}, we will analytically calculate the ratios
between critical temperature and the charge density as well as the
critical exponents of holographic p-wave superconductors. The same
procedure will be followed in Sec.\eqref{sect:swave} by studying
holographic s-wave superconductors. We will draw a conclusion in
Sec.\eqref{sect:con}.

\section{\bf Ricci Flat Gauss-Bonnet-AdS Black Holes}
\label{sect:gb}

The action of Gauss-Bonnet gravity in a $5$-dimensional  space-time
can be written as
\begin{equation}\label{action}
S=\int d^5 x \sqrt{-g} \Big(R+\frac{12}{L^2}+\frac{\alpha}{{2}}
(R^2-4R^{\mu\nu}
R_{\mu\nu}+R^{\mu\nu\rho\sigma}R_{\mu\nu\rho\sigma})\Big),
\end{equation}
where $12/L^2$ is related to the cosmological constant and $L$ is
the radius of the AdS space-time. The quadratic curvature term in
this action is the Gauss-Bonnet term with $\alpha$ the Gauss-Bonnet
coefficient.  The Ricci flat solution of this action is the
$5$-dimensional  Gauss-Bonnet-AdS black hole ~\cite{Cai:2001dz}. The
metric is described by
\begin{eqnarray}\label{metric}
ds^2
&=&-f(r)dt^2+\frac{1}{f(r)}dr^2+\frac{r^2}{L^2}(dx^2+dy^2+dz^2),
\\
f(r) &=&\frac{r^2}{2\alpha}\bigg(1-\sqrt{1-\frac{4\alpha}{L^2}(
1-\frac{ML^2}{r^4})}\bigg), \nonumber
\end{eqnarray}
where $M$ is the mass of the black hole. The horizon is located at
$r=r_+=\sqrt[4]{ML^2}$, and the temperature of the black hole is
\begin{equation}\label{temperatur}
T=\frac{r_+}{\pi L^2}.
\end{equation}
Here we should notice that in the asymptotic region with $r \to
\infty$,
\begin{equation}
f(r)\sim
\frac{r^2}{2\alpha}\Bigg[1-\sqrt{1-\frac{4\alpha}{L^2}}\Bigg].
\end{equation}
Therefore, one can define an effective radius $L_{\rm{eff}}$ of the
AdS spacetime by
\begin{equation}\label{L2}
L_{\rm{eff}}^2\equiv\frac{2\alpha}{1-\sqrt{1-\frac{4\alpha}{L^2}}}.
\end{equation}
We can see from this equation that one has to have $\alpha\leq
L^2/4$ in order to have a well-defined vacuum for the gravity
theory. The upper bound $\alpha =L^2/4$ is called Chern-Simons
limit. In the AdS/CFT correspondence, this asymptotic AdS spacetime
is dual to a conformal field theory living on the boundary
$r\rightarrow\infty$. The temperature of the black hole is just the
one of the dual field theory.  For simplicity of making integration
and making comparison to the previous numerical results, we will
assume $0<\al\leq L^2/5$ in the following context.

\section{Holographic P-wave Superconductors}
\label{sect:pwave}

\subsection{Basic setup}

In order to study the holographic p-wave superconductors in the
probe limit, we will introduce an SU(2) Yang-Mills action in to the
bulk theory, the Lagrangian density is

\begin{equation}\label{action}
\mathscr{L}_p={-}\frac{1}{4}F^a_{\mu\nu}F^{a\mu\nu},
\end{equation}
where,  $F^a_{\mu\nu}=\partial_\mu A^a_\nu-\partial_\nu A^a_\mu +
\epsilon^{abc}A^b_\mu A^c_\nu$ is the Yang-Mills field strength,
$a,b,c=(1,2,3)$ are the indices of the generators of SU(2)
 algebra. $\mu,\nu=(t,r,x,y,z)$ are the labels of space-time with
 $r$ being the radial coordinate of AdS. The $A^a_{\mu}$ are the
 components of the mixed-valued gauge
 fields
 $A=A^a_{\mu}\tau^adx^{\mu}$, where $\tau^a$ are the SU(2)
 generators with commutation
relation $[\tau^a,\tau^b]=\eps^{abc}\tau^c$.
  $\epsilon^{abc}$ is the totally antisymmetric tensor
with $\epsilon^{123}=+1$.

 Following  Refs.~\cite{Gubser:2008wv,Manvelyan:2008sv,Basu:2009vv,Ammon:2009xh,Akhavan:2010bf}, we choose the ansatz of the gauge fields as
  \be\label{ansatz} A(r)=\phi(r)\tau^3dt+\psi(r)\tau^1dx.\ee
 In this ansatz we regard the U(1) symmetry generated by $\tau^3$ as the U(1) subgroup of
 SU(2). We call this U(1) subgroup as U(1)$_3$. The
gauge boson with nonzero component $\psi(r)$ along $x$ direction is
charged under $A^3_t=\phi(r)$. According to AdS/CFT dictionary,
$\phi(r)$ is dual to the chemical potential in the boundary field
theory while $\psi(r)$ is dual to the $x$ component of some charged
vector operator $J$. The condensation of $\psi(r)$ will
spontaneously break the U(1)$_3$ gauge symmetry and induce the
phenomena of superconducting on the boundary field theory.

 The Yang-Mills equations with the above ansatz (\ref{ansatz}) are
\be\label{ym} \left\{ \begin{array}{ll} \r_r^2\phi+\frac{3
}{r}\r_r\phi -\frac{L^2 \psi^2}{r^2 f}
\phi=0,\\
\\
\r_r^2\psi +(\frac{1}{r}+\frac{\r_rf }{f})\r_r\psi +\frac{\phi ^2
}{f^2}\psi=0.
\end{array}\right.
\ee In order to solve the equations (\ref{ym}), we need the boundary
conditions for the fields $\psi(r)$ and $\phi(r)$. At the black hole
horizon, it is required that $\phi(r_+)=0$ for the U(1) gauge field
to have a finite norm, and $\psi(r_+)$ should be finite. Near the
boundary of the bulk, we have
\begin{eqnarray}
\label{bphi}\phi(r)&\rightarrow& \mu - \rho/r^2 \\
\label{bpsi}\psi(r)&\rightarrow& \psi^{(0)} + \psi^{(2)}/r^2.
\end{eqnarray}
$\mu$ and $\rho$ are dual to the chemical potential and charge
density of the boundary CFT , $\psi^{(0)}$ and $\psi^{(2)}$ are dual
to the source and expectation value of the boundary operator $J_x^1$
respectively. We always set the source $\psi^{(0)}$ to  zero, as we
want to have a normalizable solution.

\subsection{Relations between the critical temperature and charge density}

In this subsection, we will adopt S-L method, which was pioneered by
\cite{Siopsis:2010uq}, to analytically calculate the ratios between
the critical temperature $T_c$ and the charge density $\rho$ of the
holographic p-wave superconductors.

Define $z=\frac{r_+}{r}$,  the EoMs \eqref{ym} will become

 \be \label{ymz}
  \left\{\begin{array}{c}
           \phi''-\frac{\phi'}{z}-\frac{L^2\psi^2}{z^2 f}\phi=0,
           \\\\
           \psi''+(\frac1
           z+\frac{f'}{f})\psi'+\frac{\phi^2r_+^2}{f^2z^4}\psi=0.
         \end{array}  \right.
         \ee
where a prime denotes the derivative with respect to $z$. In the
following we will scale $L=1$. When the temperature approaches the
critical temperature {\it i.e.}, $T\rha T_c$, the condensation
approaches zero, viz. $\psi\rha0$, from eq.\eqref{ymz} we can get
 \be \phi''-\frac{\phi'}{z}\approx0.\ee
With the boundary conditions \eqref{bphi}, we can get that near the
critical temperature, the electric field behaves like
 \be\label{la} \phi(z)\approx \la r_+(1-z^2).\ee
 where, $\la=\rho/r_+^3$.  Near the boundary, we will introduce a
 trial function $F(z)$ into $\psi(z)$ as in \cite{Siopsis:2010uq}
  \be\label{J} \psi|_{z\rha0}\sim\psi^{(2)}\frac{z^2}{r_+^2}\sim \langle
  J_x^1\rangle\frac{z^2}{r_+^2}F(z).\ee
  where, $F(0)=1, F'(0)=0$. Therefore, using \eqref{la} and \eqref{J} the EoM of $\psi(z)$
  reduces to
  \be\label{Feom}
  F''(z)&&+\frac{3+4\al(5z^4-3)-3\sqrt{1+4\al(z^4-1)}}{\big(1+4\al(z^4-1)-\sqrt{1+4\al(z^4-1)}\big)z}F'(z)
  \nno\\&&+\frac{16\al
  z^2}{1+4\al(z^4-1)-\sqrt{1+4\al(z^4-1)}}F(z)+\la^2\frac{4\al^2(z^2-1)^2}{(\sqrt{1+4\al(z^4-1)}-1)^2}F(z)=0.\nno\\\ee
  Multiplying both sides of the above equation with
  \be T(z)=\frac{z^3}{2\sqrt \al}\big(\sqrt{1+4\al(z^4-1)}-1\big),\ee
  We will convert the eq.\eqref{Feom} to be
  \be
  \frac{d}{dz}\bigg(T(z)F'(z)\bigg)+\underbrace{\frac{8\sqrt{\al}z^5}{\sqrt{1+4\al(z^4-1)}}}_{-Q(z)}F(z)+\la^2
  \underbrace{\bigg(\frac{4\al^2(z^2-1)^2T(z)}{(\sqrt{1+4\al(z^4-1)}-1)^2}\bigg)}_{P(z)}F(z)=0.\nno\\
  \ee
 From the Sturm-Liouville eigenvlaue problem \cite{Gelfand:1963}, the
 minimum of eigenvalues of $\la^2$ can be obtained from the variation
 of the following functional
  \be \la^2[F(z)]=\frac{\int_0^1 dz\big(T(z)F'(z)^2+Q(z)F(z)^2\big)}{\int_0^1dz P(z)F(z)^2},\ee
The trial function $F(z)$ can be assumed to be
 \be F(z)\equiv1-a z^2,\ee
 which satisfies the boundary condition. So $\la^2(z)$ can be
 explicitly written as
  \be \la^2(z)=\frac{s(a,\al)}{t(a,\al)}.\ee
  where,

\be s(a,\al)&=&  60(4 \alpha -1) \left((4 \alpha
   -1) a^2+4 \alpha \right) \log \left(\frac{(1-4
   \alpha ) \alpha }{\left(2 \alpha +\sqrt{\alpha
   }\right)^2}\right) \nno\\&&-80 \sqrt{\alpha } \bigg(12
   \alpha +a \left(20 \alpha  a-3 a+32 \sqrt{1-4
   \alpha } \alpha -48 \alpha -8 \sqrt{1-4 \alpha
   }+8\right)\bigg),\nno\\
 t(a,\al)&=&\frac{15}{2} (a+1) \alpha  \left[128 (a+1) \log
   \left(\frac{-4 \alpha +\sqrt{1-4 \alpha
   }+1}{2-8 \alpha }\right) \alpha ^{3/2}\right.\nno\\&&\left.+\left(16
   (3 a+4) \alpha ^2+8 (3 a+2) \alpha -a\right)
   \log \left(\frac{(1-4 \alpha ) \alpha }{\left(2
   \alpha +\sqrt{\alpha
   }\right)^2}\right)\right]\nno\\&&-2 \sqrt{\alpha }
   \bigg(\left(\alpha  \left(32 \alpha  \left(9
   \sqrt{1-4 \alpha }-30 \log 2+5\right)+56
   \sqrt{1-4 \alpha }-45\right)-2 \sqrt{1-4 \alpha
   }+2\right) a^2\nno\\&&+5 \alpha  \left(4 \alpha
   \left(32 \sqrt{1-4 \alpha }-96 \log
   2+11\right)+16 \sqrt{1-4 \alpha }-13\right)
   a\nno\\&&+20 \alpha  \left(\alpha  \left(20 \sqrt{1-4
   \alpha }-48 \log 2+6\right)+\sqrt{1-4 \alpha
   }-1\right)\bigg)\ee

For different values of $\al$, we list the minimum value of $\la^2$
and the corresponding value of $a$ in Table\eqref{pwave}. Because
$T=r_{+}/\pi$ and $\la=\rho/r_+^3$, we can easily get that when
$T\sim T_c$, there is
 \be T_c\approx\ga ~\sqrt[3]{\rho},\ee
 where, $\ga=1/({\pi\la_{\rm min}^{1/3}})$.

 In Table\eqref{pwave}, we also list the analytical values and
 numerical values \cite{Cai:2010cv} of $\ga$ for different $\al$. The
 differences between the analytical and numerical values are within
 $1\%$. We can find that when $\al$ grows, the ratio $\ga$ will
 decrease, which means the critical temperature will get lower
 compared to the charge density. This implies that the greater the
 Gauss-Bonnet coefficient is, the harder the condensation to form.
 This statement is consistent with the previous numerical results
 \cite{Gregory:2009fj,Cai:2010cv,Pan:2009xa,Cai:2010zm}.

\begin{table}
\caption{\label{pwave} The minimum values of $\la^2$ and the
corresponding $a$ for different $\al$ are listed in the first three
columns. The analytical and numerical values of $\ga$ and $\zeta$
are shown in the last four columns. The differences between the
analytical and numerical values of $\ga$ is about $1\%$ for various
$\al$'s. The order for $\zeta_{\rm analytical}$ and $\zeta_{\rm
numerical}$ is the same.}
 \centering
\begin{tabular} {|c|c|c||c|c||c|c|}
  \hline
  $\al$ & $a$ & $\la_{\rm min}^2$ & $\ga_{\rm analytical}$ & $\ga_{\rm numerical}$ &$\zeta_{\rm analytical}$& $\zeta_{\rm numerical}$\\
  \hline\hline
  $0.0001$ & $0.685$ & $16.748$ & $0.199$ & $0.201$ &$326.858$&$499.036$\\
  \hline
  $0.1$ & $0.677$ & $20.604$ & $0.192$ & $0.194$&$337.735$&$509.796$ \\
  \hline
  $0.2$ & $0.653$ & $28.571$ & $0.182$ & $0.183$&$349.614$&$519.583$ \\
  \hline
 \end{tabular}
\end{table}

\subsection{Critical exponents and the condensation values}

From last subsection we know that when $T\rha T_c$, the condensation
value of the dual operator $\langle J_x^1\rangle$ is very small.
Using formula \eqref{J} we can convert the eq.\eqref{ymz} to be
 \be\label{phieom}\phi''-\frac{\phi'}{z}=\frac{\mathcal{A}z^2F^2}{f}\phi,\ee
where, $\mathcal{A}=\langle J_x^1\rangle^2/r_+^4$. Because
$\mathcal{A}$ is small, considering \eqref{la} we can expand
$\phi(z)$ in $\mathcal{A}$ as
 \be \frac\phi {r_+}\sim \la(1-z^2)+\mathcal{A}~\chi(z)+\cdots,\ee
 From the boundary condition we can get $\chi(1)=0$, and for simplicity we set
 $\chi'(1)=0$ \cite{Siopsis:2010uq}. The eq.\eqref{phieom} now becomes
 \be \label{chieom}\chi''-\frac{\chi'}{z}=\la\frac{z^2F^2(1-z^2)}{f}.\ee

 When temperature is away but very close to the critical
 temperature, we expand $\phi(z)$ near $z=0$,
  \be\label{expan}
  \phi=\mu-\frac\rho{r_+^2}z^2=\frac{\rho}{r_+^2}(1-z^2)\Rightarrow
  \frac\phi{r_+}&=&\frac\rho{r_+^3}(1-z^2)\sim\la(1-z^2)+\mathcal{A}\chi(z)+\cdots\nno\\
   &=&\la(1-z^2)+\mathcal{A}\big(\chi(0)+\chi'(0)z+\frac{\chi''(0)}{2}z^2+\cdots\big).\nno\\\ee
   Comparing the coefficients of $z^2$ term in the above formula~\footnote{Here, the analysis
is a little different from \cite{Siopsis:2010uq} in which they
compared the coefficients of  $z$ term. The main difference comes
from the near boundary behavior of $\phi(z)$ which deeply originates
from the dimension of space-time. } we can get
 \be \label{rho}\frac{\rho}{r_+^3}=\la-\frac {\mathcal A}{ 2}\chi''(0).\ee
From the eq.\eqref{chieom} we know that
 \be
 \chi''(0)=[\frac{\chi'(z)}{z}+\frac{\mathcal Az^2F^2}{f}\phi]\bigg|_{z\rha0}=[\frac{\chi'(z)}{z}]\bigg|_{z\rha0}\ee
Because the second term in the {\it r.h.s} of the first equality
approaches zero when $z\rha0$. In order to calculate the value of
$[\frac{\chi'(z)}{z}]\bigg|_{z\rha0}$, fortunately we find that
multiply $1/z$ to  both sides of eq.\eqref{chieom} we get
 \be(\frac1 z\chi')'=\la\frac{zF^2(1-z^2)}{f}\ee
 Integrate both sides of the above equation and note that
 $\chi'(1)=0$, we obtain~\footnote{Comparing the $z$ term in \eqref{expan}, we know that
 $\chi'(0)\rha 0$ when $z\rha 0$. This is consistent with \eqref{chiprime}.}
 \be \label{chiprime}[\frac{\chi'(z)}{z}]\bigg|_{z\rha0}=-\la\int_0^1dz
 \frac{zF^2(1-z^2)}{f}\ee
Therefore, from \eqref{rho} one has
 \be \frac{\rho}{r_+^3}=\la+\frac {\mathcal A}{ 2} \la\int_0^1dz
 \frac{zF^2(1-z^2)}{f}\ee
From the form of $\mathcal A$ and $f(z)$ we can obtain
 \be\frac{\rho}{\la r_+^3}=1+\frac{\mathcal B\langle J_x^1\rangle^2}{2
 r_+^6}\ee
where $ \mathcal B=\int_0^1dz \big(zF^2(z)(1-z^2)/g(z)\big)$ and
$g(z)=f(z)/r_+^2$. So we can deduce that
 \be \langle J_x^1\rangle=\zeta T_c^3\sqrt{1-\frac{T}{T_c}}.\ee
 where $\zeta=\sqrt6\pi^3/\sqrt{\mathcal B}$.

  The analytical values and numerical values \cite{Cai:2010cv} of
  $\zeta$ for various $\al$'s can be found in Table.\eqref{pwave}.
 They match up to the same order. Besides, in both analytical and numerical cases,
   the ratio $\zeta$ increases when $\al$ grows bigger, which implies that the condensation
   values for the dual operators will also increase. From the AdS/CFT dictionary, this
   operator can be interpreted as the operator for the paring
   mechanism, the bigger vacuum expectation value of this operator will
   make the condensation harder. This statement is also in agreement
   with the previous arguments  \cite{Gregory:2009fj,Cai:2010cv,Pan:2009xa,Cai:2010zm}.
.

 \section{Holographic S-wave Superconductors}
 \label{sect:swave}
\subsection{Basic setup}
In order to study the holographic s-wave superconductors, we
consider the Maxwell field couples to a complex scalar field in the
bulk.  The Lagrangian density is \cite{Hartnoll:2008vx},
 \be \mathscr{L}_s=-\frac1
 4F_{\mu\nu}F^{\mu\nu}-|\nabla\psi-iA\psi|^2-m^2|\psi|^2\ee
 where $A_{\mu}(r)$ is the vector field and $m^2$ is the mass square of the
 scalar field. As usual, we can choose the gauge of vector field as
 $A_{\mu}=(\phi(r),0,0,0)$ which is consistent with the ansatz of
 a real scalar field  $\psi(r)=\psi(r)^*$. In the following, we also
 scale the AdS radius to be $L=1$.

 The EoMs for $\phi(r)$ and $\psi(r)$ are
 \be\label{eom}
 \left\{\begin{array}{c}
          \r_r^2\phi+\frac3 r\r_r\phi-\frac{2\psi^2}{f}\phi=0, \\\\
          \r_r^2\psi+(\frac{\r_r f}{f}+\frac3
          r)\r_r\psi+(\frac{\phi^2}{f^2}-\frac{m^2}{f})\psi=0.
        \end{array}\right.
 \ee
Near the  boundary, $\phi(r)$ and $\psi(r)$ behave like
  \be\label{phibd} \phi(r)&=&\mu-\frac{\rho}{r^2},\\
    \label{psibd}  \psi(r)&=&\frac{\psi^-}{r^{\Dl_-}}+\frac{\psi^+}{r^{\Dl_+}}.\ee
where, $\mu$ and $\rho$ have the same interpretation as those in
Sec.\eqref{sect:pwave}. $\Dl_{\pm}=2\pm\sqrt{4+m^2L_{\rm eff}^2}$ is
the conformal dimension of the dual operator $J$ in the boundary
field theory. We will also set $\psi^-=0$ and $\langle
J\rangle=\psi^+$. Therefore, we will unambiguously define
$\Dl\equiv\Dl_+$ from now on. On the horizon, $\phi(r)=0$ because of
the well definition of the norm.

\subsection{Relations between the critical temperature and charge density}

Then, we can define $z=r_+/r$, the EoMs \eqref{eom} become
 \be\label{eomz} \left\{
 \begin{array}{c}
 \phi''-\frac1 z\phi'-\frac{2\psi^2r_+^2}{fz^4}\phi=0,\\\\
 \psi''+(\frac{f'}{f}-\frac1z)\psi'+(\frac{\phi^2}{f^2}-\frac{m^2}{f})\frac{r_+^2}{z^4}\psi=0.
 \end{array}\right.
 \ee
where a prime denotes the derivative with respect to $z$.

Because when $T\rha T_c$, the condensation is very small,
$\psi\sim0$. Therefore, the EoM of $\phi(z)$ becomes
 \be \phi''-\frac1z\phi'=0.\ee
 With the boundary condition \eqref{phibd}, we can deduce that near the critical temperature
  $\phi(z)\sim\la r_+(1-z^2)$ where $\la\equiv\rho/r_+^3$.

As in the last section, we can introduce a trial function $F(z)$
into the asymptotical behavior of $\psi(z)$,
  \be\psi\big|_{z\rha0}\sim\frac{\langle J\rangle}{r_+^{\Dl}}z^{\Dl}F(z).\ee
where $F(0)=1, F'(0)=0$.
 Now, the EoM for scalar field reduces to
 \be\label{Feoms} F''(z)+p(z)F'(z)+q(z)F(z)+\la^2 w(z) F(z)=0.\ee
 where, \be p(z)&=&\frac{\left(1-\sqrt{4 \left(z^4-1\right) \alpha
   +1}\right) (2 \Delta -3)+4 \alpha  \left((2
   \Delta -1) z^4-2 \Delta +3\right)}{z \left(4
   \left(z^4-1\right) \alpha -\sqrt{4
   \left(z^4-1\right) \alpha +1}+1\right)}\\
            q(z)&=&\frac1 z\bigg[-\frac{\left(3 \left(\sqrt{4 \left(z^4-1\right)
   \alpha +1}-1\right)-4 \left(z^4-3\right) \alpha
   \right) \Delta }{-4 \left(z^4-1\right) \alpha
   +\sqrt{4 \left(z^4-1\right) \alpha +1}-1}\nno\\&&+\frac{2 m^2 \alpha }{\sqrt{4 \left(z^4-1\right)
   \alpha +1}-1}+(\Delta -1) \Delta\bigg] \\
            w(z)&=&\frac{4 \left(z^2-1\right)^2 \alpha
   ^2}{\left(\sqrt{4 \left(z^4-1\right) \alpha
   +1}-1\right)^2}.\ee
Multiply both sides of \eqref{Feoms} with
   \be
   T(z)=\frac{z^{2\Dl-3}}{2\sqrt{\al}}(\sqrt{1+4\al
   (z^4-1)}-1),
   \ee
we obtain the following equation
   \be \bigg(T(z)F'(z)\bigg)'-Q(z)F(z)+\la^2P(z)F(z)=0\ee
   where \be Q(z)&=&-T(z)q(z)\\
             P(z)&=&T(z)w(z)\ee
Following the procedure used in the above section, we know that the
minimum value of $\la^2$ can be obtained from variation of the
following functional
 \be \la^2[F(z)]=\frac{\int_0^1 dz\big(T(z)F'(z)^2+Q(z)F(z)^2\big)}{\int_0^1dz P(z)F(z)^2},\ee
It is easy to find that $F(z)=1-az^2$ and in the calculation we will
assume $m^2=-3/L_{\rm eff}^2$ in order to make comparison with the
previous results \cite{Gregory:2009fj}. So $\la^2(z)$ can be
explicitly written as
 \be \la^2(z)=\frac{s(a,\al)}{t(a,\al)},\ee
where, \be s(a,\al)&=& -\frac{1}{768 \alpha ^2}\bigg[  2
\sqrt{\alpha } \bigg(\left(96 \sqrt{1-4 \alpha }
   \alpha +404 \alpha -75\right) a^2\nno\\&&+48 \left(14
   \sqrt{1-4 \alpha } \alpha -24 \alpha -5 \sqrt{1-4
   \alpha }+5\right) a+144 \left(2 \sqrt{1-4 \alpha }
   \alpha +\alpha \right)\bigg)\nno\\&&+3 (4 \alpha -1)
   \bigg(25 (4 \alpha -1) a^2+144 \alpha \bigg)
   \sinh ^{-1}\left(2 \sqrt{\frac{\alpha }{1-4 \alpha
   }}\right)\bigg]\\
    t(a,\al)&=&\frac{ \sqrt{1-4 \alpha }}{480\al^{3/2}}\bigg(
\left(-144 \alpha ^2-28
   \alpha +1\right) a^2-40 \alpha  (8 \alpha +1) a-10
   \alpha  (20 \alpha +1)\bigg)\nno\\
   &&+\frac{1}{2} (a+1)^2 \sqrt{\alpha } \log \left(-4
   \alpha +\sqrt{1-4 \alpha }+1\right)\nno\\&&+\frac{(a+1) \left(16 \alpha  (4 \alpha +1)+a \left(48
   \alpha ^2+24 \alpha -1\right)\right) \log
   \left(\sqrt{1-4 \alpha } \sqrt{\alpha
   }\right)}{128 \alpha }\nno\\&&+\frac{1}{1920\al^{3/2}}\bigg[-960 (a+1)^2 \log (2-8 \alpha ) \alpha ^2\nno\\&&
   -15 (a+1)
   \left(16 \alpha  (4 \alpha +1)+a \left(48 \alpha
   ^2+24 \alpha -1\right)\right) \log \left(2 \alpha
   +\sqrt{\alpha }\right) \sqrt{\alpha }\nno\\&&+\left(920 \alpha ^2+90 \alpha -4\right) a^2+10 \alpha
    (212 \alpha +13) a+960 (a+1)^2 \log (2) \alpha ^2\nno\\&&+40 (24 \alpha +1)
   \alpha-960 (a+1)^2 \alpha ^2\Ga'(\frac3
2)/\Ga(\frac3 2) -960 (a+1)^2 \gamma_{\rm Euler}  \alpha ^2\bigg]\ee
in which, $\Ga(z)$ is the {\it gamma function}
 and $\gamma_{\rm Euler}$ is the {\it Euler's
constant}. Because $T=r_{+}/\pi$ and $\la=\rho/r_+^3$, we can easily
get that when $T\sim T_c$, there is
 \be T_c=\ga ~\sqrt[3]{\rho},\ee
 where, $\ga=1/({\pi\la_{\rm min}^{1/3}})$.

We list the analytical results in Table\eqref{tab}. We can find that
the values of $\ga$ obtained  from the S-L approach are very close
to those calculated in \cite{Gregory:2009fj} by the matching
methods~\footnote{Consult  Eq.~(3.27) in
Ref.\cite{Gregory:2009fj}.}. The differences are smaller than $4\%$
for various $\al$'s. We also find that $\ga$ decreases when $\al$
grows implying that the condensation is harder to form, which is
consistent with the previous results
\cite{Gregory:2009fj,Pan:2009xa}.

\begin{table}
\caption{\label{tab} For $m^2=-3/L_{\rm eff}^2$, the minimum values
of $\la^2$ and the corresponding $a$ for various $\al$'s are listed
in the first three columns. The analytical values from S-L approach
are very similar to those obtained from matching methods
\cite{Gregory:2009fj} with differences smaller than $4\%$. The
values of $\zeta$'s are matching up to the same order.} \centering
\begin{tabular}{|c|c|c||c|c||c|c|}
  \hline
  $\al$ & $a$ & $\la_{\rm min}^2$ & $\ga_{\rm S-L}$ & $\ga_{\rm matching}$&$\zeta_{\rm S-L}$&$\zeta_{\rm matching}$ \\
  \hline\hline
  $0.0001$&$0.722$ & $12.155$ & $0.210$ & $0.202$ &$7.705$&$6.369$ \\
  \hline
  $0.1$ & $0.706$& $14.290$ & $0.204$ & $0.197$ &$7.913$&$6.884$ \\
  \hline
  $0.2$ &$0.662$& $18.091$ & $0.196$ & $0.193$&$8.042$&$7.596$  \\
  \hline
\end{tabular}
\end{table}

\subsection{Critical exponents and the condensation values}
 When $T\rha T_c$, from the equation of $\phi(z)$, we get
 \be\label{phiz2} \phi''-\frac1 z\phi'=\frac{2\mathcal A\phi z^{2\Dl-4}F^2}{g}\ee
where, $\mathcal A\equiv \langle
J\rangle^2/r_+^{2\Dl}$,$g(z)=f(z)/r_+^2$. Because near the critical
temperature $T_c$, $\mathcal A$ is small.  We can expand $\phi(z)$
in $\mathcal A$ near $z=0$,
 \be\label{phiz} \frac{\phi}{r_+}=\la(1-z^2)+\mathcal A\chi(z)+\cdots,\ee
Comparing the coefficients in $z^2$ terms, we can get from the above
formula that
 \be\label{rhos} \frac{\rho}{r_+^3}=\la-\frac {\mathcal A}{ 2}\chi''(0)\ee
Substituting \eqref{phiz} into eq.\eqref{phiz2}, we can get EoM for
$\chi(z)$ as
 \be \chi''-\frac1 z\chi'=2\la \frac{z^{2\Dl-4}F^2(1-z^2)}{g}\ee
Multiplying $1/z$ to both sides and following the procedure in last
section, we obtain
 \be \chi''(0)=\frac{\chi'(z)}{z}\bigg|_{z\rha0}=-2\la\int_0^1dz
 \frac{z^{2\Dl-5}F^2(1-z^2)}{g}.\ee
Therefore, \eqref{rhos} becomes
 \be \frac{\rho}{r_+^3}=\la(1+\frac {\mathcal A}{
2}{\mathcal B})\ee
 where $\mathcal B=2\int_0^1dz
 z^{2\Dl-5}F^2(1-z^2)/g$.

Finally, using the formula of $\mathcal A$ we get that when $T\rha
T_c$, the condensation of the operator $J$ behaves like
  \be \langle
J\rangle=\zeta (\pi T_c)^{\Dl}\sqrt{1-\frac T{T_c}}\ee
 where, $\zeta=\sqrt6 /\sqrt{\mathcal B}$. In Table\eqref{tab} we list the
 value of  $\zeta$ in S-L method and those in matching method
 \cite{Gregory:2009fj}\footnote{Refer to  Eq.~(3.28) in Ref.\cite{Gregory:2009fj} }. We find that these two kinds of values are
 matching up to the same order. The ratio $\zeta$ increases when
 $\al$ grows bigger, which implies that the condensation will be
 harder to form. This is consistent with the previous results~\cite{Gregory:2009fj,Pan:2009xa}.

\section{Conclusions}
\label{sect:con}

In this paper, we have used the S-L method to analytically calculate
the holographic p-wave and s-wave superconductors with Gauss-Bonnet
gravity in the probe limit. In particular, we calculate the ratios
$\ga$ between the critical temperature and the charge density on the
boundary. The ratios are in great agreement with the previous
numerical results or those obtained from the analytical matching
methods. We find that the ratio will decrease when the Gauss-Bonnet
couplings $\al$ increases, which implies that the bigger $\al$ is
the harder the condensation to form. This statement is in agreement
with the previous conclusions. Besides, we also calculate the ratio
$\zeta$ between the condensation value and the critical temperature.
Although the values of ratio $\zeta$ is not so well agreement with
the numerical results or those from matching methods, they are
consistent up to the same order. Moreover, we find that when $\al$
increases, the ratio $\zeta$ will also increase. This again implies
that the bigger $\al$ is the harder the condensation to form, which
also agrees with the previous conclusions. The universal critical
exponents $1/2$ in mean-field theory was also analytically confirmed
in both p-wave and s-wave case of Gauss-Bonnet gravity. This
critical exponents will not be affected by the dimension of the
space-time or the Gauss-Bonnet coefficients $\al$. This value $1/2$
is an exact property of the Lagrangian for the matter field of
systems, and has nothing to do with the gravitational
backgrounds~\cite{Liu}.

\acknowledgments
HFL and HQZ would like to thank Xin Gao, Bin Hu, Zhang-Yu Nie and
Yuan-Jiang Zhang for their helpful discussions and comments. HFL
would be very grateful for the hospitalities of the members in the
Institute of Theoretical Physics, Chinese Academy of Sciences. This
work was supported in part by the National Natural Science
Foundation of China (No. 10821504, No. 10975168, No.11035008 and
No.11075098), and in part by the Ministry of Science and Technology
of China under Grant No. 2010CB833004.


\end{document}